\documentclass[12pt]{article}

\usepackage{wider}
\begin{document}

\newcommand{\sqvb}{\ensuremath{ \langle \!\langle 0 |} }
\newcommand{\sqvk}{\ensuremath{ | 0 \rangle \!\rangle } }
\newcommand{\sqvn}{\ensuremath{ \langle \! \langle 0 |  0 \rangle \! \rangle} }
\newcommand{\wh}{\ensuremath{\widehat}}
\newcommand{\be}{\begin{equation}}
\newcommand{\ee}{\end{equation}}
\newcommand{\bea}{\begin{eqnarray}}
\newcommand{\eea}{\end{eqnarray}}
\newcommand{\ra}{\ensuremath{\rangle}}
\newcommand{\la}{\ensuremath{\langle}}
\newcommand{\rra}{\ensuremath{ \rangle \! \rangle }}
\newcommand{\lla}{\ensuremath{ \langle \! \langle }}
\newcommand{\str}{\rule[-.125cm]{0cm}{.5cm}}
\newcommand{\pr}{\ensuremath{^\prime}}
\newcommand{\ppr}{\ensuremath{^{\prime \prime}}}
\newcommand{\da}{\ensuremath{^\dag}}
\newcommand{\as}{^\ast}
\newcommand{\eps}{\ensuremath{\epsilon}}
\newcommand{\ve}{\ensuremath{\vec}}
\newcommand{\ka}{\kappa}
\newcommand{\non}{\ensuremath{\nonumber}}
\newcommand{\lf}{\ensuremath{\left}}
\newcommand{\rt}{\ensuremath{\right}}
\newcommand{\al}{\ensuremath{\alpha}}
\newcommand{\dfn}{\ensuremath{\equiv}}
\newcommand{\ga}{\ensuremath{\gamma}}
\newcommand{\ti}{\ensuremath{\tilde}}
\newcommand{\wti}{\ensuremath{\widetilde}}
\newcommand{\hs}{\ensuremath{\hspace*{.5cm}}}
\newcommand{\bet}{\ensuremath{\beta}}

\newcommand{\cO}{\ensuremath{{\cal O}}}
\newcommand{\cS}{\ensuremath{{\cal S}}}
\newcommand{\cF}{\ensuremath{{\cal F}}}

\newcommand{\pup}{\ensuremath{^{(p)}}}

\title{\bf There Is No Basis Ambiguity in Everett Quantum Mechanics\thanks{
This work was sponsored by the Air Force under Air Force Contract 
F19628-00-C-0002.  Opinions, interpretations, conclusions, and 
recommendations are those of the author and are not necessarily endorsed 
by the U.S. Government.
}
}
\author{
Mark A. Rubin\\
\mbox{}\\ 
Lincoln Laboratory\\ 
Massachusetts Institute of Technology\\ 
244 Wood Street\\                       
Lexington, Massachusetts 02420-9185\\      
rubin@LL.mit.edu\\ 
}
\date{\mbox{}}

\maketitle

\begin{abstract}
The Everett-interpretation description of isolated measurements, i.e., measurements involving interaction between  a measuring apparatus and a measured system but not interaction with the environment, is shown to be unambiguous, claims in the literature to the contrary notwithstanding.  The appearance of ambiguity in such measurements is engendered by the fact that,  in the Schr\"{o}dinger picture, information on splitting into Everett copies must be inferred from the history of the combined system. In the Heisenberg picture this information is contained in mathematical quantities associated with a single time. 

\end{abstract}

\section{\bf Introduction} \label{Introduction}

The explanation of the quantum-mechanical measurement process  provided by the Everett interpretation 
[1-4]  
has been said to suffer from a difficulty which has been  termed the ``basis ambiguity,''  
the ``preferred basis problem,''  
or simply the ``basis problem.''  
The claim is   that the  Everett description of  measurement  is  ambiguous when the measurement involves  a measuring apparatus and measured system  which are completely isolated from interaction with any external environment   
[5-10].
According to Zurek [5], it is possible in the case of such an isolated measurement situation that
``the apparatus by virtue of being correlated with the state of the system contains not only all the information about the observable $\wh{S}$\/ [which it was designed to measure]; it must equally well contain all the information about many other observables $\wh{R}$\/ \ldots defined on the Hilbert space of the system \ldots This is so despite the fact that $\wh{R}$\/ and $\wh{S}$\/ do not, in general, commute\ldots 
Quantum mechanics alone, when applied to an isolated, composite object consisting of an apparatus and a system, cannot in principle determine which observable has been measured.''

As  emphasized by DeWitt [11], Everett quantum mechanics does in fact contain an  unambiguous notion of measurement, even in the case of  an apparatus and a system isolated from the environment.  ``The quantum mechanical description of the measurement starts with a state in which the  system and apparatus, in the absence of coupling, would be {\em uncorrelated},\/ which simply means that the state vector of the combination is expressible as a tensor product \ldots The structure of the coupling operator \ldots defines a {\em preferred}\/ set of basis vectors \ldots [11]'' (See also [12].)
Below, I show explicitly that a measurement situation defined along these lines presents no  ambiguity regarding  ``which observable has been measured.''

  Use of the basis defined by the interaction may be objected to on the grounds that ``to give to the measurement process a privileged position over other interactions  seems contrary to the spirit of Everett's program, which was motivated in part by a reaction against the privileged status of measurement (reduction of the state vector) in the orthodox interpretation [13].''  The ``privileged status'' of  measurements in Everett quantum mechanics is, however, only an artifact of the Schr\"{o}dinger picture. In the Everett interpretation in the Schr\"{o}dinger picture, measurement interactions define a decomposition of the  state vector  into  ``branches,'' each of which is considered to correspond to a globally-defined world [14] in which observers perceive definite outcomes to the measurement.
Thus in Everett quantum mechanics in the Schr\"{o}dinger picture, measurement interactions do have   the  appearance of exerting  special nonlocal influences which other types of interactions do not exert.

In Heisenberg-picture Everett quantum mechanics, all dynamics, including the dynamics of measurements, is described by the evolution of operators which represent the properties of local physical systems [15].  During a measurement interaction the operator representing that degree of freedom of the measuring device relevant for its role as measuring device is changed into a characteristic form (see Sec. \ref{HeisenbergPicture}).  This form serves to indicate the presence, subsequent to the measurement,  of several noninteracting ``Everett copies'' of that degree of freedom, each corresponding to a different outcome of the measurement [16, 17].  Operators corresponding to other degrees of freedom  are not affected, except to the extent that they subsequently interact with the copied degree of freedom.\footnote{The ontology of the Everett interpretation in the Heisenberg picture is thus different from that in the Schr\"{o}dinger picture.  It is this difference which allows for the emergence of Born-rule-consistent probability, in the familiar sense of ``probability as long term relative frequency,'' in Heisenberg-picture Everett quantum mechanics [17].}
Measuring devices are thus not ``privileged,'' merely ``different,'' in that  they interact in such a way  as to lead to Everett copying.  This copying is an explicitly local process, assuming of course that the underlying Hamiltonian is local. 

There is another drawback to using the Schr\"{o}dinger picture to describe measurement, one which is the cause of the apparent ambiguity present in isolated measurements.  The Schr\"{o}dinger picture in a sense {\em hides}\/ information about the splitting into Everett copies as a result of measurement, in that this information is not present in  mathematical quantities associated with a given time, but, rather,  must be inferred from  the history of the system up to that time (see Sec. \ref{SchrodingerPicture}).  In the Heisenberg picture, on the other hand, the manner in which systems have been split into Everett copies at time $t$\/ can be determined from mathematical quantities defined at time $t$\/ (see Sec. \ref{HeisenbergPicture}).

Sec. \ref{MeasurementSituations} below describes the conditions for the existence of a measurement situation. Sec. \ref{SchrodingerPicture} discusses Everett copying and the description of measurements in the Schr\"{o}dinger picture.  The (apparent) basis ambiguity is presented, as well as a proof of the (actual) uniqueness of the measurement basis. The hidden nature of Everett-copying information in the Schr\"{o}dinger picture is also discussed. Sec. \ref{HeisenbergPicture} presents the (less familiar) Heisenberg-picture description of the Everett interpretation, uniqueness theorems for Everett copying and  measurement  in this picture, and a proof that a measuring apparatus cannot simultaneously measure noncommuting observables. Sec. \ref{Discussion} discusses the role of interaction with the environment in Everett-quantum-mechanical measurement.

\section{\bf Measurement Situations} \label{MeasurementSituations}

We  deal in this paper only with ideal measurements (see, e.g., [18, Sec. 14.2]). 
Consider a physical system the Hilbert space of which is spanned by $M$\/ basis vectors,
\be
|\cS:i\ra,\hspace*{.5cm} i=1,\ldots,M, \label{Sstatesdef}
\ee
satisfying
\be
\la \cS:i|\cS:j \ra = \delta_{ij},\hspace*{.5cm} i,j=1,\ldots,M. \label{Sstates}
\ee
This  physical system, which is the one to be measured, will be referred to below  as ``the system'' or ${\cS}$\/.\\

\noindent {\bf Conditions for the existence of a measurement situation:}
\begin{itemize}
\item[M1.] There is present, in addition to $\cS$\/,  another physical system,
a measuring apparatus which we will term ``the observer'' or $\cO$\/. The Hilbert space of $\cO$\/ is spanned by $M+1$\/  basis vectors,
\be
|\cO:i\ra,\hspace*{.5cm} i=0,\ldots,M, \label{Obasis}
\ee
\be
\la \cO:i|\cO:j \ra = \delta_{ij},\hspace*{.5cm} i,j=0,\ldots,M. \label{Obasisortho}
\ee
The state $|\cO:0\ra$\/ is termed the ``ignorant state'' or the ``ready state.'' 
\item[M2.] Between an initial time $t_{in}$\/ and a  later time $t>t_{in}$\/,  $\cS$\/ and  $\cO$\/ evolve under the action of a unitary operator $\wh{U}$\/ which has the property that
\be
\wh{U}|\cO:0\ra |\cS:i\ra= |\cO:i\ra |\cS:i\ra, \hspace*{.5cm}i=1,\ldots,M. \label{Uaction}
\ee
\item[M3.] At  time $t_{in}$\/  the state vector of $\cS$ and  $\cO$\/ is of the form
\be
|\psi(t_{in})\ra=|\cO:0\ra |\cS;\psi\ra.  \label{psi_in}
\ee
That is, $\cO$\/ is in the ignorant state, and $\cS$\/ is in an arbitrary state
\be
|\cS;\psi\ra = \sum_{i=1}^M \psi_i |\cS:i\ra, \label{psi_S_in}
\ee
where
\be
\sum_{i=1}^M |\psi_i|^2=1, \hspace{.5cm}  {\hbox{\rm $\{\psi_i\}$\/ otherwise arbitrary,}} \label{psiarb}
\ee  
\end{itemize}

When conditions M1-M3  are satisfied, a measurement of $\cS$\/ in the $|\cS:i\ra$\/ basis is considered to have been made by $\cO$\/ at time $t$\/.

\section{\bf Schr\"{o}dinger Picture} \label{SchrodingerPicture}

\subsection{Everett Copies} \label{EverettCopies}

In the Schr\"{o}dinger picture, the time-dependent state vector is given, at time $t$\/, by 
\be
|\psi(t)\ra=\wh{U}|\psi(t_{in})\ra. \label{Sevol}
\ee
By (\ref{Uaction})-(\ref{psi_S_in}) and (\ref{Sevol}),  
\be
|\psi(t)\ra=%\wh{U}|\psi(t_{in})\ra=
\sum_{i=1}^M \psi_i |\cO:i\ra |\cS:i\ra. \label{psi_t}
\ee

So, in the Schr\"{o}dinger picture we can if we wish replace the condition M3 above for the existence of a measurement with the following condition:
\begin{itemize}
\item[M3$\mbox{}.\pr$\/]At time $t$\/ the state vector of $\cS$ and  $\cO$\/ is of the form
(\ref{psi_t}), 
where (\ref{psiarb}) is satisfied.
\end{itemize}
This is equivalent to M3 since, 
using (\ref{Uaction}),   it  follows from (\ref{psi_t}) that the state vector at the earlier time
$t_{in}$\/ is of the form (\ref{psi_in}). \\

\noindent{\bf Condition for the existence of Everett copies (Schr\"{o}dinger picture):}
In a measurement situation, each term of the  time-$t$\/  state vector  (\ref{psi_t})
represents a distinct physical reality in which $\cO$\/ has measured $\cS$\/ to be in the $i^{th}$\/ of the basis states  (\ref{Sstatesdef}).

\subsection{The ``Basis Ambiguity''} \label{TheBasisAmbiguity}

Ambiguity is introduced if we adopt the view that a measurement is characterized completely by the fact that the state vector has attained the form (\ref{psi_t}) in some basis; i.e., we pay attention to condition M3$\mbox{}\pr$\/ but disregard condition M2 for the existence of a measurement situation. 

Suppose, for example, that M=2 and that $\psi_1=\psi_2=1/\sqrt 2$\/:
\be
|\psi(t)\ra=(1/\sqrt 2)( |\cO:1\ra |\cS:1\ra + |\cO:2\ra |\cS:2\ra). \label{psi_2_t}
\ee
If we define new bases for the respective state spaces of $\cS$\/  and $\cO$\/,
\bea 
|\cS\pr :1 \ra&=&(1/\sqrt 2)(|\cS :1 \ra + |\cS : 2\ra), \label{psi_pr_1}\\
|\cS\pr :2 \ra&=&(1/\sqrt 2)(|\cS :1 \ra - |\cS : 2\ra), \label{psi_pr_2}
\eea
and
\bea
|\cO\pr:0\ra&=&|\cO:0\ra, \label{psi_O_pr_0}\\
|\cO\pr:1\ra&=&(1/\sqrt 2)(|\cO :1 \ra + |\cO : 2\ra), \label{psi_O_pr_1}\\
|\cO\pr:2\ra&=&(1/\sqrt 2)(|\cO :1 \ra - |\cO : 2\ra), \label{psi_O_pr_2}
\eea
then we can rewrite (\ref{psi_2_t}) as 
\be
|\psi(t)\ra=(1/\sqrt 2)( |\cO\pr :1\ra |\cS\pr :1\ra + |\cO\pr :2\ra |\cS\pr:2\ra )\label{psi_2_t_pr}
\ee
Comparing (\ref{psi_2_t}) and 
(\ref{psi_2_t_pr}), we conclude that the form of the decomposition of $|\psi(t)\ra$\/ is {\em by itself}\/
insufficient to unambiguously determine the nature of the splitting into Everett copies.  

If the  state vector (\ref{psi_2_t_pr}) were to  describe two Everett worlds in which respective observer copies in states $|\cO:\pr 1 \ra,$\/ $|\cO:\pr 2 \ra$\/ are correlated with system states $|\cS:\pr 1 \ra,$\/ $|\cS:\pr 2 \ra$,\/  condition M2 would have to be satisfied. But, if the states  $|\cO: 1 \ra,$\/ $|\cO: 2 \ra$,\/  $|\cS: 1 \ra$\/ and $|\cS: 2 \ra$\/ satisfy condition M2, then the states $|\cO:\pr 1 \ra,$\/ $|\cO:\pr 2 \ra$,\/  $|\cS:\pr 1 \ra$\/ and $|\cS:\pr 2 \ra $\/ do not. That is,  if
\be
\wh{U}|\cO:0\ra |\cS:1\ra= |\cO:1\ra |\cS:1\ra
\ee
and
\be
\wh{U}|\cO:0\ra |\cS:2\ra= |\cO:2\ra |\cS:2\ra,
\ee
it follows from (\ref{psi_pr_1})-(\ref{psi_O_pr_2}) that
\bea
\wh{U}|\cO :0\ra |\cS\pr :1\ra &=& (1/\sqrt 2)(|\cO\pr:1\ra |\cS\pr:1\ra + |\cO\pr:2\ra |\cS\pr:2\ra) \nonumber \\
&\not = & |\cO\pr:1\ra |\cS\pr:1\ra \label{nogo1}
\eea
and
\bea
\wh{U}|\cO :0\ra |\cS\pr :2\ra &=& (1/\sqrt 2)(|\cO\pr:1\ra |\cS\pr:2\ra + |\cO\pr:2\ra |\cS\pr:1\ra) \nonumber \\
&\not = & |\cO\pr:2\ra |\cS\pr:2\ra \label{nogo2}
\eea

The results (\ref{nogo1}) and (\ref{nogo2}) above illustrate the \\

\noindent{\bf Isolated measurement uniqueness theorem (Schr\"{o}dinger picture):} Let $|\cS:i\ra,$\/ $i=1,\ldots, M,$\/ and   $|\cO:i\ra $,\/ $i=0,\ldots,M,$\/ be  orthonormal bases for the state spaces of $\cS$\/ and $\cO$\/, respectively, and  let them satisfy (\ref{Uaction}). Then any other orthonormal bases $|\cS\pr:i\ra,$\/ $i=1,\ldots, M,$\/  and  $|\cO\pr:i\ra $,\/ $i=0,\ldots,M,$\/ of $\cS$\/ and $\cO$\/ for which
\be
|\cO\pr:0\ra=|\cO:0\ra
\ee
and which also  satisfy (\ref{Uaction}), i.e.,
\be
\wh{U}|\cO : 0\ra |\cS \pr : i \ra= |\cO\pr:i\ra |\cS\pr:i\ra, \hspace*{.5cm}i=1,\ldots,M, \label{Uaction_pr}
\ee
are essentially identical to $|\cS : i\ra$\/ and   $|\cO : i \ra $\/, in that
\be 
|\cS\pr :i\ra=a_i |\cS : \pi(i)\ra, \hspace*{.5cm}i=1,\ldots M,
\ee
and
\be
|\cO\pr :i \ra =|\cO :\pi(i)\ra,  \hspace*{.5cm}i=1,\ldots,M,
\ee
where $\pi(i)$ is some permutation of $i=1,\ldots,M,$\/ and $|a_i|=1$\/, $i=1,\ldots M$\/.\\

\noindent {\bf Proof:} Expand $|\cS\pr : i\ra$\/ in terms of $|\cS : i\ra$\/:
\be
|\cS\pr : i\ra = \sum_{j=1}^M a_{ij}|\cS : j\ra, \hspace*{.5cm}i=1,\ldots,M, \label{psi_pr_exp}
\ee
so
\be
\la\cS:j|\cS\pr:i\ra=a_{ij}. \label{psi_psi_pr_inner_prod}
\ee
using (\ref{Sstates}).
Using (\ref{Uaction}) and (\ref{psi_pr_exp}),
\be
\wh{U} |\cO:0\ra|\cS\pr:i\ra=\sum_{j=1}^Ma_{ij} |\cO:j\ra |\cS:j\ra. \label{Uaction_inter}
\ee
Using (\ref{Uaction_pr}) with (\ref{Uaction_inter}),
\be
\sum_{j=1}^Ma_{ij} |\cO:j\ra |\cS:j\ra=|\cO\pr:i\ra |\cS\pr:i\ra. \label{Uaction_inter_2}
\ee
Taking the inner product of (\ref{Uaction_inter_2}) with the bra $\la \cO: l| \la \cS: l|$\/ and using (\ref{psi_psi_pr_inner_prod}) and the orthonormality of the unprimed bases,
\be
a_{il}(1-\la \cO:l| \cO\pr : i\ra)=0, \hspace*{0.5cm}i,l=1,\ldots,M. \label{master_eq_sch}
\ee

Fix $i=i_1$\/. There must exist at least one value of $l$\/, say $l_1$\/, such that $a_{i_1l_1}\not =
0$\/ (else, from  (\ref{psi_pr_exp}), $|\cS\pr:i_1\ra$\/ would have zero norm, contradicting the assumed orthonormality of the primed basis for $\cS$\/.) Then, from (\ref{master_eq_sch}),
\be
\la \cO : l_1|\cO\pr:i_1\ra=1. \label{O_Opr_inner_prod_11}
\ee
Since $|\cO\pr : i_1\ra$\/  has unit norm and the $|\cO : l\ra$\/  form a basis,
\bea
1 & = & \la \cO\pr:i_1 |  \cO\pr:i_1 \ra  \nonumber \\
  &= &\sum_{l=0}^M \la \cO\pr:i_1 | \cO  : l \ra \la \cO: l |\cO\pr:i_1 \ra  \nonumber \\
  &=&1+\sum_{l \not = l_1} |\la \cO : l | \cO\pr : i_1\ra |^2
\eea
using (\ref{O_Opr_inner_prod_11}), which implies that $\la \cO : l | \cO\pr : i_1\ra =0$\/
for $l \not = l_1$\/ or, with (\ref{O_Opr_inner_prod_11}),
\be
\la \cO : l | \cO\pr : i_1\ra =\delta_{ll_1},  \hspace*{0.5cm}l=1,\ldots,M. \label{O_Opr_inner_prod_1}
\ee
Using (\ref{master_eq_sch}) with (\ref{O_Opr_inner_prod_1}),
\be
a_{i_1 l}=a_{i_1}\delta_{l l_1},  \hspace*{0.5cm}l=1,\ldots,M. \label{a_i1_l}
\ee

For $i=i_2 \not = i_1$\/, we similarly find from (\ref{master_eq_sch}) that
\be
\la \cO : l | \cO\pr : i_2\ra =\delta_{ll_2},   \hspace*{0.5cm}l=1,\ldots,M, \label{O_Opr_inner_prod_2}
\ee
and
\be
a_{i_2 l}=a_{i_2}\delta_{l l_2},   \hspace*{0.5cm}l=1,\ldots,M. \label{a_i2_l}
\ee
for some $l=l_2$\/. It is not possible that $l_2=l_1$\/. That would imply, from
(\ref{O_Opr_inner_prod_2}),
\be
\la \cO : l | \cO\pr : i_2\ra =\delta_{ll_1},   \hspace*{0.5cm}l=1,\ldots,M, \label{O_Opr_inner_prod_wrong_12}
\ee
or, setting $l=l_1$\/,
\be
\la \cO : l_1 | \cO\pr : i_2\ra =1. \label{O_Opr_inner_prod_wrong_12b}
\ee
From (\ref{O_Opr_inner_prod_1}) with $l=l_1$\/,
\be
\la\cO:l_1|\cO\pr:i_1\ra=1.\label{O_l1_O_i2_norm}
\ee
Since $|\cO:l_1\ra$\/ is normalized and the $|\cO\pr:i\ra$\/'s form a basis, we obtain  
a contradiction:
\bea
1&=&\la \cO:l_1|\cO:l_1\ra  \nonumber \\
 &=&\sum_{i=0}^M \la \cO:l_1|\cO\pr : i\ra \la \cO\pr:i |\cO:l_1\ra \nonumber \\
 &=&1 + 1 + \sum_{i \not = i_1, i_2} |\la \cO : l_1|\cO \pr : i \ra |^2
\eea
using (\ref{O_Opr_inner_prod_wrong_12b}) and (\ref{O_l1_O_i2_norm}), implying 
$1 \ge 2$\/.

Continuing in this manner for $i=i_3,\ldots,i_M,$\/ we obtain a mapping $i_1 \rightarrow  l_1$\/, \ldots, $i_M \rightarrow  l_M$\/ which is some permutation $\pi(i)$\/ of the integers $i=1, \ldots, M$\/.  In terms of $\pi(i)$\/,  
(\ref{O_Opr_inner_prod_1})-(\ref{a_i2_l})  
generalize to 
\be
a_{il}=a_i\delta_{l\pi(i)},   \hspace*{0.5cm}i,l=1,\ldots,M. \label{a_il_final}
\ee
and
\be
\la \cO : l | \cO\pr : i \ra =\delta_{l\pi(i)},  \hspace*{0.5cm}i,l=1,\ldots,M. 
\label{O_O_pr_inner_prod}
\ee
Using (\ref{a_il_final}) in (\ref{psi_pr_exp}),
\be
|\cS\pr: i\ra=a_i |\cS:\pi(i)\ra, \hspace*{0.5cm}i=1,\ldots,M.
\ee
So,
\be
|a_i|=1,\hspace*{0.5cm}i=1,\ldots,M,
\ee
since $|\cS\pr: i\ra$\/ and $|\cS:\pi(i)\ra$\/ are both normalized.
From (\ref{O_O_pr_inner_prod}) and the orthonormality of the primed and unprimed bases of the state space of $\cO$\/,
\be
|\cO\pr: i\ra=|\cO:\pi(i)\ra, \hspace*{0.5cm} i=1,\ldots,M.
\ee
{\bf Q.E.D.}\\

The conclusion of the  above theorem would of course be changed if we permitted the use of  a different evolution operator $\wh{U}\pr \not = \wh{U}$\/ or a different ready state $|\cO\pr:0\ra \not = |\cO:0\ra$\/ in the condition (\ref{Uaction_pr}) for $|\cS\pr:i\ra$\/ and $|\cO\pr:i\ra$\/.
However, this would correspond physically to utilizing a  different measuring apparatus, and would have no bearing on the question of whether or not the Everett splitting produced by measurement with a particular measuring apparatus is ambiguous.

\subsection{Hidden Information in the Schr\"{o}dinger Picture} \label{HiddenInformationintheSchrodingerPicture}

There is thus no basis ambiguity in Everett-quantum-mechanical measurements. There is, however, a peculiarity, to say the least, in the Schr\"{o}dinger-picture description of such measurements, specifically in the  characterization of splitting into Everett copies.   We are used to thinking of the Schr\"{o}dinger picture state vector at a given time $t$\/, $|\psi(t)\ra$\/, as containing all the information there is to know about the physical situation at time $t$\/. However, we see that the state vector by itself does not contain all the information needed to know how the combined system of $\cS$\/ and $\cO$\/ has split into correlated Everett copies. We also need to know the past history of $\cS$\/ and $\cO$\/, i.e., that they underwent the time evolution generated by $\wh{U}$\/.  In the case that the  corresponding Hamiltonian is time-independent, one might argue that the information describing splitting at time $t$\/ is present  in mathematical quantities characterizing the combined system at time $t$\/, namely $ |\psi(t)\ra$\/ together with the Hamiltonian. In the case of a time-dependent Hamiltonian, one could not even make this argument.

Deutsch and Hayden [15], in analyzing the spatial location of quantum information, point out that ``in the Schr\"{o}dinger picture \ldots it is impossible to characterize quantum information at a given instant using the state vector alone. To investigate where information is located, one must also take into account how that state came about.'' Here we see that knowing how the Schr\"{o}dinger-picture state came about is also needed to characterize  
Everett-copying information. 
This peculiarity is {\em not}\/ present in the Heisenberg-picture representation of the spatial location of information [15]---nor is it present in the Heisenberg-picture representation of Everett-copying  information, as we will now show.

\section{\bf Heisenberg Picture} \label{HeisenbergPicture}

\subsection{Operators and Dynamics} \label{OperatorsandDynamics}

The  conditions for existence of a measurement situation are the same as in Sec. 
\ref{MeasurementSituations}. The state spaces (\ref{Sstatesdef}), (\ref{Obasis}) are also the same, so condition M1 is satisfied.
We  take the time-independent Heisenberg-picture state vector to be $|\psi(t_{in})\ra$ as given in (\ref{psi_in}),  so condition M3  is satisfied.

Heisenberg-picture operators are time-dependent. For any operator $\wh{o}$\/,
\be
\wh{o}(t)=\wh{U}\da \wh{o}(t_{in}) \wh{U}. \label{UactionHP}
\ee
Where there is no possibility of confusion we will drop the time argument for operators evaluated at the initial time $t_{in}$:
\be
\wh{o}=\wh{o}(t_{in}).
\ee
Let $\wh{a}$\/ and $\wh{b}$\/ be  nondegenerate operators in the respective state spaces of
$\cS$\/ and $\cO$\/ which have the basis vectors (\ref{Sstatesdef}) and (\ref{Obasis})
as eigenvectors:
\be
\wh{a}|\cS;\al_i\ra=\al_i|\cS;\al_i\ra, \hspace*{.5cm}i=1,\ldots,M, \label{HPbasisS}
\ee
\be
\al_i = \al_j \Rightarrow i=j, \hspace*{.5cm}i,j=1,\ldots,M,\label{anondegen}
\ee
\be
\wh{b}|\cO;\bet_i\ra=\bet_i|\cO;\bet_i\ra, \hspace*{.5cm}i=0,\ldots,M,\label{HPbasisO}
\ee
\be
\bet_i = \bet_j \Rightarrow i=j, \hspace*{.5cm}i,j=0,\ldots,M,\label{bnondegen}
\ee
where
\be
|\cS;\al_i\ra \equiv |\cS: i\ra, \hspace*{.5cm}i=1,\ldots,M, \label{newSnotation}
\ee
\be
|\cO;\bet_i\ra \equiv |\cO: i\ra, \hspace*{.5cm}i=0,\ldots,M. \label{newOnotation}
\ee

To obtain explicit expressions for operators at time $t$\/, we need an explicit expression for $\wh{U}$\/.  The time evolution operator $\wh{U}$\/ is generated by a Hamiltonian $\wh{H}$\/,
\be
\wh{U}=\exp(-i\wh{H}(t-t_{in})). \label{U_iHt}
\ee
For ideal measurements, $\wh{H}$\/ has the form (see, e.g, [19])
\be
\wh{H}=\sum_{i=1}^M \wh{h}_i^{\cO} \otimes \wh{P}_i^{\cS}, \label{H_u_P}
\ee
where
\be
\wh{P}_i^{\cS}= |\cS;\al_i\ra\la \cS;\al_i|, \label{PS_def}
\ee
so
\be
\wh{P}_i^{\cS}\wh{P}_j^{\cS}=\delta_{ij}\wh{P}_j^{\cS} \label{PP_delta}
\ee
and 
\be
\sum_{i=1}^M \wh{P}_i^{\cS}=1, \label{PS_completeness}
\ee
and where $\wh{h}_i^{\cO}$\/ acts nontrivially only in the state space of $\cO$\/.  
Using this fact and (\ref{U_iHt})-(\ref{PS_completeness}),
\be
\wh{U}=\sum_{i=1}^M\wh{u}_i^{\cO} \otimes \wh{P}_i^{\cS}, \label{Uform}
\ee
where
\be
\wh{u}_i^{\cO}= \exp(-i \wh{h}_i^{\cO}(t-t_{in})).  \label{uform}
\ee
If the $\wh{u}_i^{\cO}$\/ satisfy
\be
\wh{u}_i^{\cO}|\cO;\bet_0\ra=|\cO;\bet_i\ra, \hspace*{.5cm}i=1,\ldots,M, \label{orthou}
\ee
then, using 
(\ref{Sstates}), (\ref{newSnotation}), (\ref{newOnotation}), (\ref{PS_def}) and (\ref{orthou}),
we see that $\wh{U}$\/ in (\ref{Uform}) satisfies (\ref{Uaction}), so condition M2 for a measurement situation is satisfied.\footnote{In general a $\wh{U}$\/ satisfying (\ref{Uaction}) could have extra terms which annihilate $|\cO;\bet_0\ra$\/, in addition to those present on the right hand side of (\ref{Uform}). As we see here, such terms will be absent if $\wh{U}\/$ is generated by a Hamiltonian, and that Hamiltonian is of the form (\ref{H_u_P}), i.e., one which does not disturb the states $|\cS;\al_i\ra$\/.}  (For example, 
\be
\wh{h}_i^{\cO}=i\kappa (|\cO;\bet_i\ra\la \cO;\bet_0|-|\cO;\bet_0\ra\la \cO;\bet_i|)
\label{particular_h}
\ee 
leads via (\ref{uform}) to $\wh{u}_i^{\cO}$\/'s satisfying (\ref{orthou}),
provided
\be
\kappa=\frac{\pi}{2(t-t_{in})}.)
\ee

Using (\ref{Uform}) it follows that any operator $\wh{d}$\/ which at time $t_{in}$\/ acts nontrivially only in the state space of $\cO$\/  has, at time $t$\/,  the form
\be
\wh{d}(t)=\sum_{i=1}^M \wh{d}_i \otimes \wh{P}_i^{\cS}, \label{dform}
\ee
where
\be
 \wh{d}_i=\wh{u}_i^{\cO\dagger}\wh{d}\wh{u}_i^{\cO}.
\ee
In particular,  
\be
\wh{b}(t)=\sum_{i=1}^M \wh{b}_i \otimes \wh{P}_i^{\cS}, \label{bsplitform}
\ee
where
\be
 \wh{b}_i=\wh{u}_i^{\cO\dagger}\wh{b}\wh{u}_i^{\cO}. \label{bidef}
\ee
From (\ref{HPbasisO}), (\ref{orthou}) and (\ref{bidef})
\be
\wh{b}_i|\cO;\bet_0\ra=\bet_i|\cO;\bet_0\ra.\label{biaction}
\ee

\subsection{Interpretation} \label{Interpretation}

\subsubsection{Everett Copies} \label{EverettCopiesHP}

\noindent {\bf Condition for the existence of Everett copies (Heisenberg Picture):}\footnote{This condition is termed ``interpretive rule 1'' in [17].
``Interpretive rule 2'' of [17], relating to probability, will not be required in the present paper.}
If an operator $\wh{ b}(t)$\/ can be expressed in the form (\ref{bsplitform}), where  
(\ref{biaction}) is satisfied for nondegenerate $\beta_i$\/ and the constant Heisenberg-picture state vector is of the product form (\ref{psi_in}), then  
that operator is considered to represent  
$M$\/ Everett copies.\footnote{One may well inquire at this juncture, ``Everett copies of {\em what}\/?'' When this condition arises in the context of a measurement situation, the answer is 
``Everett copies of the degree of freedom of $\cO$\/ represented at time $t_{in}$\/ by $\wh{b}$\/.'' (See Sec. \ref{Measurement}.) In this context we conclude, from (\ref{psi_in}) and from  (\ref{HPbasisO}) with $i=0$\/, that $\wh{b}$\/ has the value $\bet_0$\/, indicating ignorance ($\wh{b}|\psi(t_{in})\ra = \bet_0 |\psi(t_{in})\ra $\/), while, from  (\ref{psi_in}) and (\ref{biaction}), we see that  the $\wh{b}_i$\/'s have the respective values $\bet_i$\/ corresponding to the possible outcomes of the measurement ($\wh{b}_i|\psi(t_{in})\ra = \bet_i |\psi(t_{in})\ra $\/).  Whether there exist conditions different from those for measurement situations which give rise to Everett copies as defined by this condition is an open question. The physical meaning of $\wh{b}(t)$\/ might also conceivably be determined by the way it enters into the Hamiltonian (the time-$t$\/ Hamiltonian, in the case of a time-dependent Hamiltonian) and interacts with operators representing other degrees of freedom.}\\

This condition  makes reference only to mathematical quantities defined at one moment of time; indeed, neither condition M2 nor anything else about the form of the time evolution operator $\wh{U}$\/ is invoked.
In the Heisenberg picture, as opposed to the Schr\"{o}dinger picture,
quantities characterizing the combined system at a single time---the time-dependent operators and the time-independent state vector--- contain all the information about the nature of splitting into Everett copies.\footnote{Information contained in the time-independent state vector can be transfered to the time-dependent operators by a unitary transformation [15, 20].}  Of course the  would-be characterization of splitting in the Schr\"{o}dinger picture  by considering only  the decomposition (\ref{psi_t}) of the state vector also refers only to quantities defined at a single time; but, as we have seen,  this characterization of splitting is ambiguous. The characterization of splitting in terms of  Heisenberg-picture operators described in the  condition above is, however, completely unambiguous, due to the\\

\noindent {\bf Operator expansion uniqueness theorem:} Let $\wh{B}$\/ be an 
operator which can be expanded as
\be
\wh{B}=\sum_{i=1}^M \wh{b}_i \pr\otimes \wh{P}_i^{\cS\pr} \label{bprform}
\ee
where each $\wh{b}_i\pr$\/, $i=1,\ldots,M$\/ acts nontrivially only in the $M+1$\/-dimensional state space of $\cO$\/ and satisfies
\be
\wh{b}_i\pr|\cO;\bet_0\pr\ra=\bet_i\pr|\cO;\bet_0\pr\ra, \hspace*{0.5cm} i=1,\ldots,M, \label{bpraction}
\ee
\be
\bet_i\pr = \bet_j\pr \Rightarrow i=j,  \hspace*{0.5cm} i,j=1,\ldots,M,\label{bprinondegen}
\ee
for some vector $|\cO;\bet_0\pr\ra$\/ in the state space of $\cO$\/,
and where 
\be
\wh{P}_i^{\cS\pr}=|\cS\pr;\al_i\pr\ra\la\cS\pr;\al_i\pr|, \hspace*{0.5cm} i=1,\ldots,M,\/ \label{PSpr_def}
\ee
with $|\cS\pr;\al_i\pr\ra$\/, $i=1,\ldots,M,$\/ orthonormal basis vectors for 
 the $M$\/-dimensional state space of $\cS$\/,
\be
\la\cS\pr;\al_i\pr|\cS\pr;\al_j\pr\ra=\delta_{ij}, \hspace*{0.5cm} i,j=1,\ldots,M, \label{Sprortho}
\ee
\be
\al_i\pr = \al_j\pr \Rightarrow i=j, \label{aprinondegen}
\ee
\be
\sum_{i=1}^M \wh{P}_i^{\cS\pr}=1. \label{SPprcomplete}
\ee
Then any other expansion of $\wh{B}$\/ satisfying these conditions---i.e., 
\be
\wh{B}=\sum_{i=1}^M \wh{b}_i \ppr\otimes \wh{P}_i^{\cS\ppr}, \label{bpprform}
\ee
where each $\wh{b}_i\ppr$\/, $i=1,\ldots,M$\/ acts nontrivially only in the  state space of $\cO$\/ and satisfies
\be
\wh{b}_i\ppr|\cO;\bet_0\pr\ra=\bet_i\ppr|\cO;\bet_0\pr\ra, \hspace*{0.5cm} i=1,\ldots,M, \label{bppraction}
\ee
\be
\bet_i\ppr = \bet_j\ppr \Rightarrow i=j, \label{bpprinondegen}
\ee
for the same vector $|\cO;\bet_0\pr\ra$,
and where 
\be
\wh{P}_i^{\cS\ppr}=|\cS\ppr;\al_i\ppr\ra\la\cS\ppr;\al_i\ppr|, \hspace*{0.5cm} i=1,\ldots,M,\/ \label{PSppr_def}
\ee
with $|\cS\ppr;\al_i\pr\ra$\/, $i=1,\ldots,M$\/ orthonormal basis vectors for 
 the state space of $\cS$\/,
\be
\la\cS\ppr;\al_i\ppr|\cS\ppr;\al_j\ppr\ra=\delta_{ij}, \hspace*{0.5cm} i,j=1,\ldots,M, \label{Spprortho}
\ee
\be
\al_i\ppr = \al_j\ppr \Rightarrow i=j, \label{apprinondegen}
\ee
\be
\sum_{i=1}^M \wh{P}_i^{\cS\ppr}=1 \label{SPpprcomplete}
\ee
---is essentially identical to (\ref{bprform}), in that
\be
\wh{P}_i^{\cS\ppr}=\wh{P}_{\pi(i)}^{\cS\pr}, \hspace*{0.5cm} i=1,\ldots,M,\label{PSprmapping}
\ee
and
\be
\wh{b}_i\ppr=\wh{b}_{\pi(i)}\pr\hspace*{0.5cm} i=1,\ldots,M,\label{bprmapping}
\ee
where $\pi(i)$\/ is some permutation of $i=1,\ldots,M$\/.\\

\noindent {\bf Proof:} From (\ref{bprform}) and (\ref{bpprform}),
\be
\sum_{i=1}^M \wh{b}_i \pr\otimes \wh{P}_i^{\cS\pr}=\sum_{i=1}^M \wh{b}_i \ppr\otimes \wh{P}_i^{\cS\ppr}.  \label{bpr_bppr}
\ee
Taking the matrix element of both sides of (\ref{bpr_bppr}) between the ket $\la \cO;\bet_0\pr| \la \cS;\al_k\pr|$\/ and the bra $| \cO;\bet_0\pr\ra |\cS;\al_j\ppr\ra$\/ yields, using 
(\ref{bpraction}), (\ref{PSpr_def}), (\ref{Sprortho}), (\ref{bppraction}), (\ref{PSppr_def}) and
(\ref{Spprortho}),
\be
(\bet_k\pr-\bet_j\ppr)\la\cS\pr;\al_k\pr|\cS\ppr;\al_j\ppr\ra=0, \hspace*{0.5cm} j,k=1,\ldots,M. \label{master_eq}
\ee
Pick some $j$,\/ call it $j_1$.\/ There must be at least one $k$\/ such that  
$\la\cS\pr;\al_k\pr|\cS\ppr;\al_{j_1}\ppr\ra \not = 0$\/,  
since $|\cS\ppr;\al_j\ppr\ra$\/ has unit norm and the 
$|\cS\pr;\al_j\pr\ra$\/'s form a basis. Pick one of these $k$\/'s, call it $k_1$\/. 
Then
\be
\la\cS\pr;\al_{k_1}\pr|\cS\ppr;\al_{j_1}\ppr\ra \not = 0,   \label{Saprk_Sapprj}
\ee
implying, with (\ref{master_eq}), 
\be
\bet_{k_1}\pr=\bet_{j_1}\ppr. \label{bprk1_bpprj1}
\ee
Furthermore, there is no other value of $k$\/, $\wti{k} \not = k_1$\/, such that 
$\la\cS\pr;\al_{\wti{k}}\pr|\cS;\al_{j_1}\ppr\ra \not = 0$\/. (If there were,  that would imply, 
with (\ref{master_eq}), that $\bet_{\wti{k}}\pr=\bet_{j_1}\ppr$\/, which with (\ref{bprk1_bpprj1}) would imply $\bet_{\wti{k}}\pr=\bet_{k_1}\pr$\/, contradicting
(\ref{bprinondegen}).) So,
\be
\la \cS\pr; \al_k\pr |\cS\ppr;\al_{j_1}\ppr\ra=a_{j_1}\delta_{k_1k}, \hspace*{0.5cm}a_{j_1} \not = 0, \hspace*{0.5cm} k=1,\ldots,M. \label{SprSpprinnerprod1}
\ee

Similarly, for $j=j_2 \not = j_1$\/, there exists a unique $k_2$\/ such that
\be
\bet_{k_2}\pr=\bet_{j_2}\ppr \label{bprk2_bpprj2}
\ee
and
\be
\la \cS\pr; \al_k\pr |\cS\ppr;\al_{j_2}\ppr\ra=a_{j_2}\delta_{k_2k}, \hspace*{0.5cm}a_{j_2} \not = 0, \hspace*{0.5cm} k=1,\ldots,M. \label{SprSpprinnerprod2}
\ee
It's not possible that $k_2 =k_1$\/, since that, with (\ref{bprk1_bpprj1}) and (\ref{bprk2_bpprj2}), would imply $\bet_{j_1}\ppr=\bet_{j_2}\ppr$\/ even though
$j_1 \not = j_2$\/, contradicting (\ref{bpprinondegen}).  

Continuing in this manner we find that (\ref{bprk1_bpprj1})-(\ref{SprSpprinnerprod2}) generalize to
\be
\bet_{\pi(j)}\pr=\bet_{j}\ppr, \hspace*{0.5cm}j=1,\ldots,M, \label{bprk_bpprj}
\ee
and
\be
\la \cS\pr; \al_k\pr |\cS\ppr;\al_{j}\ppr\ra=a_{j}\delta_{\pi(j) k}, \hspace*{0.5cm}a_{j} \not = 0, \hspace*{0.5cm} j,k=1,\ldots,M, \label{SprSpprinnerprod}
\ee
where $\pi(j)$\/ is a permutation of $j=1,\ldots,M$\/. Since $|\cS\ppr;\al_{j}\ppr\ra$\/ has unit norm and the $|\cS\pr;\al_{j}\pr\ra$\/'s form a basis, 
\be
|a_i|=1, \hspace*{0.5cm}i=1,\ldots,M. \label{anorm}
\ee
Using (\ref{PSpr_def}), (\ref{Sprortho}), (\ref{PSppr_def}), (\ref{Spprortho}), (\ref{SprSpprinnerprod}) and (\ref{anorm}), we obtain
\be
\wh{P}_i^{\cS\ppr}=\wh{P}_{\pi(i)}^{\cS\pr}, \hspace*{0.5cm} i=1,\ldots,M.\label{bprmappingagain}
\ee

Using (\ref{bprform}), (\ref{bpprform}) and (\ref{bprmappingagain}),
\be
\sum_{i=1}^M \wh{b}_i \pr\otimes \wh{P}_i^{\cS\pr} = \sum_{i=1}^M \wh{b}_i \ppr\otimes \wh{P}_{\pi(i)}^{\cS\pr} \label{bpr_e_bppr}
\ee
or
\be
\sum_{i=1}^M(\wh{b}_i\pr-\wh{b}_{\pi^{-1}(i)}\ppr) \otimes \wh{P}_i^{\cS\pr}=0. \label{bpr_m_bppr}
\ee
Let $|\cO;\chi\ra$\/ and $|\cO;\eta\ra$\/ be arbitrary vectors in the state space of $\cO$\/.  Taking the matrix element of (\ref{bpr_m_bppr}) between $\la \cO;\chi|\la \cS\pr; \al_i|$\/ and 
$| \cO;\eta\ra| \cS\pr; \al_i\ra$\/ and using (\ref{PSpr_def}) and (\ref{Sprortho}),
\be
\la \cO;\chi|(\wh{b}_i\pr-\wh{b}_{\pi^{-1}(i)}\ppr)|\cO;\eta\ra=0. \label{bpr_m_bppr_mat_elt}
\ee
Since $|\cO;\chi\ra$\/ and $|\cO;\eta\ra$\/ are arbitrary,
\be
\wh{b}_i\pr=\wh{b}_{\pi^{-1}(i)}\ppr,
\ee
or
\be
\wh{b}_i\ppr=\wh{b}_{\pi(i)}\pr.
\ee

\noindent {\bf Q.E.D.}\\

Note that, like the condition for the existence of Everett copies in the Heisenberg picture, the operator expansion uniqueness theorem  make reference only to mathematical quantities associated with a single time, and   neither invokes condition M2 nor makes any other reference to the form of the time evolution operator $\wh{U}$\/. The  theorem also makes no reference to the state vector $|\psi(t_{in})\ra$\/; it is purely a statement about operators at a single time, independent of any dynamics.

\subsubsection{Measurement} \label{Measurement}

In the Heisenberg picture we must impose an additional condition on the existence of a measurement situation, to assure that there is a suitable degree of freedom to record the measurement being made:

\begin{itemize}
\item[M4] At the initial time $t_{in}$\/ there is an operator $\wh{b}$\/ which acts nontrivially only in the state space of $\cO$\/ and which satisfies (\ref{HPbasisO}) and (\ref{bnondegen}), where the $\cO$\/-space basis vectors appearing in (\ref{HPbasisO}) are the same as those 
which appear in condition M2. %\ref{orthou}). 
\end{itemize}

Conditions M2 and M4 refer to both $t$\/ and $t_{in}$\/, so the Heisenberg-picture characterization of measurement, as distinct from that of Everett 
copying, 
 depends on quantities associated with more than a single time, as is the case in the Schr\"{o}dinger picture. 
(In the Schr\"{o}dinger picture the  condition for splitting into Everett copies is not distinct from, and in fact refers to, the conditions for the existence of a measurement situation. See Sec. \ref{EverettCopies}.) 
Any characterization of measurement must in principle refer to states of affairs before and after the measurement interaction, since which   physical quantity, if any, the measuring apparatus has recorded depends on both the initial and final states of the apparatus. (The multimeter reads ``50;'' but was it set to record voltage, or current?)

We now immediately obtain the\\

\noindent {\bf Isolated measurement uniqueness theorem (Heisenberg picture):}
Given a measurement situation, i.e., one  satisfying conditions M1-M4,  
with $\wh{U}$\/ of the form (\ref{Uform}),  
$\wh{b}(t)$\/ is split into  Everett copies in a uniquely-defined manner.\\

\noindent {\bf Proof:} We have already shown, in Sec. \ref{OperatorsandDynamics}, that $\wh{b}(t)$\/ satisfying the conditions of the theorem will be of the form (\ref{bsplitform}) with (\ref{biaction}) holding,  and (\ref{psi_in}) holds since M3 holds.   %(\ref{biaction2}) 
So the  condition for the existence of Everett copies is satisfied; and the operator expansion uniqueness theorem  shows that this splitting into copies is unambiguous.\\

\noindent {\bf Q.E.D.}

\subsubsection{Simultaneous Measurement of Noncommuting Observables?} 
\label{SimultaneousMeasurementsofNoncommutingObservables?}

Is it possible that a measuring device designed to measure an observable $\wh{a}$\/ of 
$\cS$\/ also simultaneously measures an observable $\wh{a}\pr$\/ of $\cS$\/ which does not commute with $\wh{a}$\/? That would require the existence of an operator $\wh{d}$\/ satisfying M4 which, under the action of $\wh{U}$\/, evolves to $\wh{d}(t)=\sum_{i=1}^M \wh{d}_i\pr \otimes \wh{P}_i^{\cS\pr},$\/ where the $\wh{d}_i\pr$\/'s  act in $\cO$\/ space, as well as the existence of   an $\cS$\/-space operator $\wh{a}\pr$\/ diagonal in the basis defined by the $\wh{P}_i^{\cS\pr}$\/'s and  not commuting with $\wh{a}$\/ (which by (\ref{HPbasisS}) and (56) is  diagonal in the basis defined by the $\wh{P}_i^{\cS}$\/'s). But any operator which acts only in $\cO$\/-space at time $t_{in}$\/ will be of the form (\ref{dform}) at time $t$\/ (whether or not it satisfies the rest of condition M4). By the operator expansion uniqueness  theorem,  the $\wh{P}_i^{\cS\pr}$\/'s  must be the same as the 
$\wh{P}_i^{\cS}$\/'s up to  renumbering, so $\wh{a}\pr$\/ and  $\wh{a}$\/ must commute.
It is thus not possible for a measuring device to simultaneously  measure noncommuting observables.

\section{\bf Discussion} \label{Discussion}

None of the above  implies that interaction with the environment does not play an important role in the phenomenology of  measurement.
Interaction with the environment in the form of decoherence is believed to be responsible for localization of macroscopic objects, at least to the extent that  localization is equivalent to the suppression of quantum interference [21] (see also [22] and references therein).  Recent experiments provide quantitative support for this belief [23, 24].

Besides, isolated measurements divorced from interaction with the environment are by far the exception rather than the rule, to the regret of designers of  quantum computers [25]. Including such interactions along with measurement interactions will require modification to the formalism presented above. Not all isolated measurements 
are expected to retain the property of repeatability in the presence of these modifications, only those for which the $\wh{P}_i^{\cS}$\/'s define a basis (``pointer basis'') compatible with the environmental interactions [5, 8].

However, it is one thing to say that interactions with the environment affect and constrain measurements, and another to say that interaction with the environment is a necessary condition for the very existence of well-defined measurements. 
Stapp [7] argues that interaction with the environment cannot do both the basis-selection and localization jobs at once; i.e., that if interaction with the environment is required for the selection of a preferred basis, then the Everett interpretation  either must  be at odds  with the observed fact that  
macroscopic objects are always found in spatially well-localized states or must  suffer from the basis ambiguity.  If interaction with the environment  is relieved of the task of selecting a preferred basis, it may be possible to obviate this objection to the Everett interpretation.\footnote{Stapp's  argument involves the presence of continuous spatial degrees of freedom in addition to the discrete degrees of freedom $\cS$\/ and $\cO$\/ considered here.  Measurement situations including spatial degrees of freedom will be examined elsewhere.}

\section*{Acknowledgments} 

I would like to thank Jian-Bin Mao and Allen J. Tino for helpful
discussions.

\end{document}